\documentstyle[prabib,aps,psfig,12pt]{revtex}
\newtheorem{thm}{Assumption}

\begin{document}
\title{Wormhole with Quantum Throat}
\author{V. Dzhunushaliev
\thanks{E-Mail Address : dzhun@rz.uni-potsdam.de and
dzhun@freenet.bishkek.su}}
\address{Universit\"at Potsdam, Institute f\"ur Mathematik,
14469, Potsdam, Germany \\
and Theor. Phys. Dept. KSNU, 720024, Bishkek, Kyrgyzstan}

\maketitle

\begin{abstract}
A wormhole with a quantum throat on the basis of an approximate
model of the spacetime foam is presented. An effective spinor
field is introduced for the description of the spacetime foam. The
consequences of such model of the wormhole is preventing a
``naked'' singularity in the Reissner-Nordstr\"om solution with
$|e|/m > 1$.
\end{abstract}

\section{Introduction}

\begin{figure}
\centerline{ \framebox{
\psfig{figure=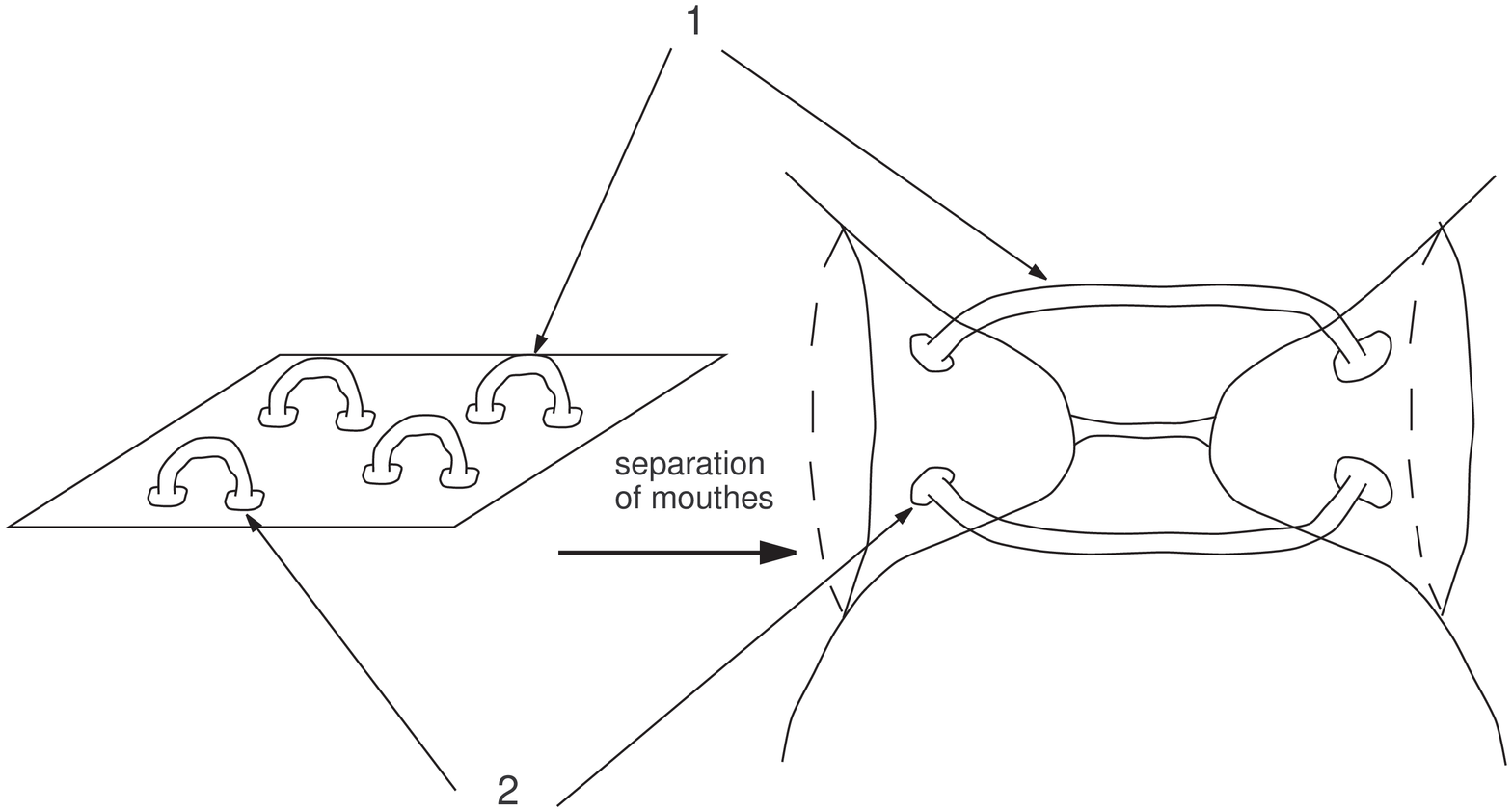,height=5cm,width=10cm,angle=-90}}}
 \vspace{5mm}
\caption{The spacetime foam with separated mouthes of quantum
handles. {\bf 1} are the quantum (virtual) handles (WHs); {\bf 2 }
are the mouthes of WH's.} \label{fig1}
\end{figure}

By definition a wormhole (WH) is a bridge connecting two
asymptotically flat regions. Usually such construction is a
classical object and should satisfy to Einstein equations (see
\cite{visser} for more detailed introduction to this field). The
topology of the 4D WH is $R_1\times R_2\times S^2$, where $R_1$ is
the time, $R_2$ is the radial coordinate and $S^2$ is the cross
section of the WH. A space region near to the minimal cross
section $S^2_{min}$ is called as a throat. The basic problem for
existing of the WH is concentrated on the throat: in this region a
matter violates the so-called null energy condition. In this paper
we offer a model of the WH in which the throat is \textit{a set of
quantum handles (wormholes) in the presence of the strong electric
field}. These quantum handles can be considered as quantum WH's of
a spacetime foam with separated mouthes (see Fig.\ref{fig1}). We
remind that the hypothesized spacetime foam \cite{wheel0},
\cite{wheel1} is a set of quantum wormholes (handles) appearing in
the spacetime on the Planck scale level ($l_{Pl} \approx
10^{-33}cm$). For the macroscopic observer these quantum
fluctuations are smoothed and we have an ordinary smooth manifold
with the metric submitting to Einstein equations.
The presence of the spacetime foam with virtual wormholes (VWH)
which have separated mouthes gives a physical effect: VWH's can
entrap a part of the electric flux lines and in spite of the fact
that at the infinity $|e|/m > 1$ we will have a non-singular
spacetime ($e$ and $m$ are the charge and mass registered at the
infinity).
\par
The mechanism for creating the WH with quantum throat can be
offered by the following way:
\begin{enumerate}
  \item
  The VWH's connect two regions with the strong electric field by
such a way that the electric flux lines leak through VWH's from
one region to another one \cite{dzh7,dzh99pb}, see Fig.\ref{fig2}.
  \item
  Each VWH is the solutions of the 5D Einstein equations with
  $G_{5t} \neq 0$.
  \item
  As usually the $G_{5t}$ metric component can be considered as the 4D
  electric field which joins the electric flux lines of the
  above-mentioned two 4D regions.
  \item
  For this model the whole spacetime is 5D one and outside
of the throat the $G_{55}$ is non-dynamical variable and we have
the 5D Kaluza-Klein theory in its initial interpretation ($G_{55}
= 1$ and 5D gravity is equivalent to Einstein-Maxwell theory),
inside of the throat the $G_{55}$ is dynamical variable and we
have the ordinary Kaluza-Klein gravity with 4D metric +
electromagnetic and scalar fields. Splitting off the 5D dimension
takes place on the event horizon. The matching conditions of the
4D and 5D fields on the event horizon is discussed in
Ref.\cite{dzh99yb}. As this composite WH is the classical solution
of the 5D Kaluza-Klein equations a probability for the quantum
creation of such VWH is not zero.
\end{enumerate}

\section{Qualitative description of the model}

\begin{figure}
\centerline{ \framebox{
\psfig{figure=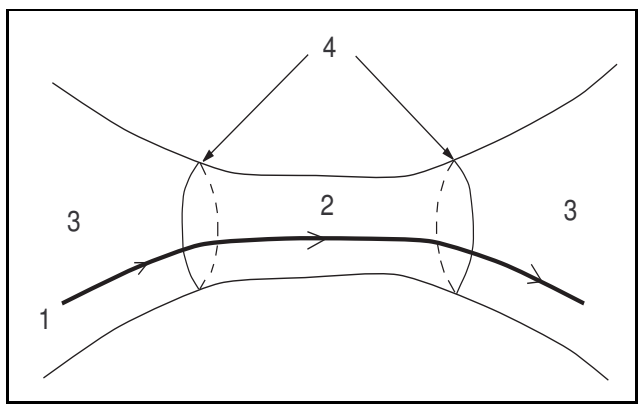,height=5cm,width=8cm}}} \vspace{5mm}
\caption{The model of the VWH in the presence of an external
electric field. {\bf 1} is the force line of the external electric
field; {\bf 2} is the 5D classical throat; {\bf 3} is the external
spacetimes; {\bf 4} are the event horizons (mouthes of the VWH).}
\label{fig2}
\end{figure}

The model of the VWH is presented on the Fig.(\ref{fig2}).
\par
\begin{figure}
\centerline{ \framebox{
\psfig{figure=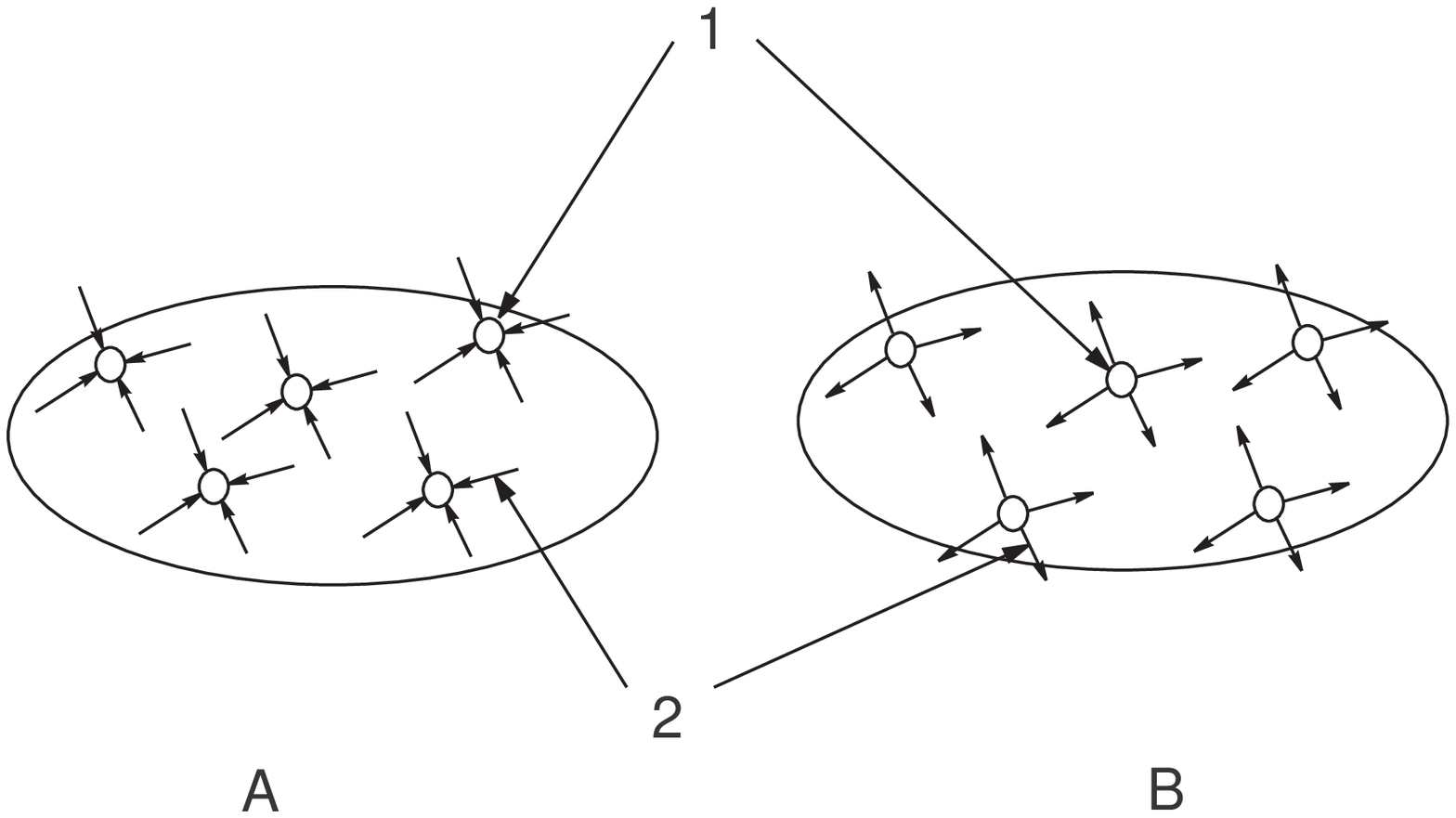,height=5cm,width=10cm}}} \vspace{5mm}
\caption{{\bf A} is the part of the WH with force lines of the
electric field incoming into mouthes of the VWH, respectively {\bf
B} is the part of the WH with force lines of the electric field
outcoming from mouthes of the VWH. {\bf 1} are the mouthes of
VWHs; {\bf 2} are the force lines of the electric field. Each
mouth has a random velocity.} \label{fig3}
\end{figure}

On the Fig.(\ref{fig3}) are presented the two parts of the WH with
the quantum throat.
On the Fig.\ref{fig3}A is presented the part of the WH with incoming
force lines of the electric field and respectively on the
Fig.\ref{fig3}B the part with outcoming force lines. For the 4D observer
each mouth is like to moving (+/-) electric charge. But this
movement is stochastic one and as a consequence we have
$\langle magnetic$ $field\rangle \approx 0$ where $\langle \rangle$
is the stochastical averaging.
\par
In Ref.\cite{vdschmidt2} is shown that such composite WH has a spin-like
structure. The cause of this is following. The metric
describing the 5D throat is
\begin{eqnarray}
ds^{2} & = & \eta_{AB}\omega ^A\omega ^B =
\nonumber \\
&& - \frac{r_0^2}{\Delta (r)}(d\chi  -
\omega (r)dt)^2 + \Delta (r)dt^{2} - dr^{2} - a(r)
\left (
d\theta ^2 + \sin\theta ^2 d\varphi ^2
\right ),
\label{ql1}\\
a & = & r^{2}_{0} + r^{2},
\quad
\Delta = \pm \frac{2r_0}{q}\frac{r^2 + r_0^2}
{r^2 - r_0^2} ,
\quad
\omega = \pm \frac{4r_0^2}{q}\frac{r}
{r^2 - r_0^2} .
\label{ql2}
\end{eqnarray}
where $\chi $ is the 5$^{th}$ extra coordinate;
$\eta_{AB} = (\pm,-,-,-,\mp)$, $A,B = 0,1,2,3,5$;
$r,\theta ,\varphi $ are the $3D$  polar coordinates;
here $r_0 > 0$ and $q$ are some constants.
We see that the signs of the $\eta_{55}$ and $\eta_{00}$
are not defined. We remark that this 5D metric is located behind
the event horizon therefore the 4D observer is not able to
determine the signs of the $\eta_{55}$ and $\eta_{00}$ \cite{vdschmidt2}.
Moreover this 5D metric (\ref{ql1}) can fluctuate between these
two possibilities. Hence the external 4D observer is forced
to describe such composite WH by means of a spinor.
\par
This allows us to do the following assumption:
\begin{thm}
The stochastical polarized spacetime foam approximately can be
determinated by some effective field: a spinor field
$\psi$.
\end{thm}
In this case a density $\epsilon$ of the VWH is
\begin{equation}
\epsilon = |\psi|^2 = \tilde\psi \psi
\label{ql5}
\end{equation}
where ($\tilde{}$) means the transposition
and a density of an effective electric charge $\rho$
(the mouthes of VWH entrapped the electric force lines) is
\begin{equation}
\rho = e\tilde \psi \psi
\label{ql6}
\end{equation}
where $e$ is some charge.

\section{Exact description of the model}

As the hypothesized spacetime foam is a consequence of quantum
gravity all our fields (metric $g_{\mu\nu}$, electromagnetic filed
$A_\mu$ and spinor field $\psi$) in this model should be quantized
fields. The interaction between these fields is very strong and we
can not use the Feynmann diagram technique. For the quantization
of this model we will use the Heisenberg quantization method which
he applied for the non-linear spinor field \cite{heis}. The
essence of this method consists in that the classical fields
$f(x^\mu)$ in field equations are excanched on the field operators
$f(x^\mu) \to \hat f(x^\mu)$. In this case we have differential
equations for operators. Certainly it is not clear what is a
solution of such differential equations. In fact Heisenberg has
shown that differential equations for the operator field is
equivalent to some infinite set of differential equations for
Green functions (for the small coupling constant this is
Dayson-Schwinger equations system).
\par
Following this way we write differential equations for the
gravitational + electromagnetic fields in the presence of the
spacetime foam $(\psi)$ as follows
\begin{eqnarray}
\hat R_{\mu\nu} -\frac{1}{2}\hat g_{\mu\nu}\hat R& = & \hat
T_{\mu\nu} ,
\label{ex1}\\
\left ( i \hat\gamma^\mu \partial _\mu
+ e\hat A_\mu - \frac{i}{4}\hat\omega _{\bar a\bar
b\mu}\hat\gamma^\mu \hat\gamma^{[\bar a}\hat\gamma^{\bar b]} - m
\right )\hat\psi & = & 0 , \label{ex2}\\ D_\nu \hat F^{\mu\nu} & =
& 4\pi e \left ( \hat{\bar\psi}\hat\gamma^\mu \hat \psi \right )
\label{ex3}
\end{eqnarray}
where $\hat R_{\mu\nu}$ is the operator of Ricci tensor;
$\hat T_{\mu\nu}$ is the operator of the sum of the energy-momentum
tensor the spinor and Maxwell fields;
$\hat g_{\mu\nu}$ is the operator of the gravitational field;
$\hat\gamma^\mu = \hat h^\mu_{\bar a}\gamma^{\bar a}$ is the
operator of the Dirac matrices;
$\hat g^{\mu\nu} = (1/2)(\hat\gamma^\mu\hat\gamma^\nu +
\hat\gamma^\nu\hat\gamma^\mu)$, $[]$ means the antisymmetrization;
$\gamma^{\bar a}$ ($\bar a$ is the vier-bein index)
is the usual $\gamma$-matrices
\begin{equation}
\gamma^{\bar 0} =
\left (
\begin{array}{cc}
1 & 0
\\
0 & 1
\\
\end{array}
 \right ) ; \quad
\gamma^{\bar i} =
\left (
\begin{array}{cc}
0 & \sigma^i
\\
-\sigma^i & 0
\\
\end{array}
 \right ) ; \quad i=1,2,3
\label{ex4}
\end{equation}
where $\sigma^i$ are the Pauli matrices; $\hat h^\mu_{\bar a}$
is the vier-bein operator; Greek indexes are the spacetime indexes;
Latin indexes with the bar are vier-bein indexes;
$\hat F_{\mu\nu} = \partial_\mu \hat A_\nu - \partial_\nu \hat A_\mu$
is the operator of the Maxwell tensor of the electromagnetic field;
$\hat A_\mu$
is the operator of the potential; $m$ and $e$
are some constants. Certainly this equation system is hopelessly
complicated, and are impossible to find exact solutions.
\par
We will consider equations for average values
$\langle g_{\mu\nu}\rangle$,
$\langle A_\mu \rangle$, $\langle \psi\rangle$ and so on.
In the first approximation we suppose that
\begin{equation}
\langle \hat f(x)\rangle \approx f(\langle \hat x\rangle) ,
\label{ex5}
\end{equation}
where $\hat f$ can be $\hat R_{\mu\nu}$,
$\hat \omega_{\bar a \bar b\mu}$ and so on; $\hat x$ can be
$\hat g_{\mu\nu}$, $\hat\psi$, $\hat A_\mu$ and so on.
In this case we have the classical system of
the Einstein-Dirac-Maxwell equations, {\it i.e.} now
the system (\ref{ex1})-(\ref{ex3}) is without ($\hat{}$).
\par
\begin{figure}
\centerline{ \framebox{
\psfig{figure=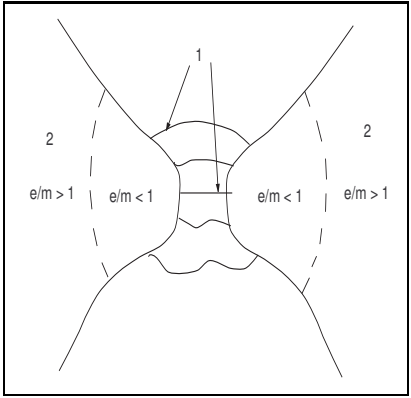,height=5cm,width=5cm}}} \vspace{5mm}
\caption{{\bf 1} are the quantum (virtual) WHs, {\bf 2} are two
solutions with $|e_\infty|/m_\infty>1$. Such object can be named
as \textit{the wormhole with quantum throat}.} \label{fig4}
\end{figure}

Fot our model we use the following ansatz:
for the spherically symmetric metric
\begin{equation}
ds^2 = e^{2\nu(r)}\Delta (r) dt^2 -
\frac{dr^2}{\Delta (r)} - r^2
\left (
d\theta ^2 + \sin ^2d\varphi ^2
\right ) ,
\label{ex6}
\end{equation}
for the electromagnetic potential
\begin{equation}
A_\mu =
\left (
-\phi,0,0,0
\right ) ,
\label{ex7}
\end{equation}
for the spinor field
\begin{equation}
\tilde \psi = e^{-i\omega t}\frac{e^{-\nu /2}}{r\Delta ^{1/4}}
\left (
f,0,ig\cos\theta ,ig\sin\theta e^{i\varphi}
\right ) .
\label{ex8}
\end{equation}
The following is \textit{very important} for us: the ansatz
(\ref{ex8}) for the spinor field $\psi$ has the $T_{t\varphi}$
component of the energy-momentum tensor and the $J^\varphi = 4\pi
e (\bar\psi \gamma^\varphi \psi)$ component of the current. Let we
remind that $\psi$ determines the stochastical gas of the VWH's
which can not have a preferred direction in the spacetime. This
means that after substitution expression (\ref{ex6})-(\ref{ex8})
into field equations they should be averaged by the spin direction
of the ansatz (\ref{ex8})\footnote{another words the averaging
$\langle \rangle$ in the expression (\ref{ex5}) is not only
quantum but stochastical, too.}. After this averaging we have
$T_{t\varphi} = 0$ and $J^\varphi = 0$ and we have the following
equations system describing our spherically symmetric spacetime
\begin{eqnarray}
f' \sqrt{\Delta} & = & \frac{f}{r} - g
\left (
  \left (
  \omega - e\phi
  \right )\frac{e^{-\nu}}{\sqrt\Delta} + m
\right ) ,
\label{ex9}\\
g' \sqrt{\Delta} & = & f
\left (
  \left (
  \omega - e\phi
  \right )\frac{e^{-\nu}}{\sqrt\Delta} - m
\right ) -
\frac{g}{r} ,
\label{ex10}\\
r\Delta ' & = & 1 - \Delta -
\kappa \frac{e^{-2\nu}}{\Delta}
\left (\omega - e\phi \right )
\left (f^2 + g^2 \right ) - r^2e^{-2\nu}{\phi'}^2
\label{ex11a}\\
r\Delta\nu ' & = & \kappa\frac{e^{-2\nu}}{\Delta}
\left (\omega - e\phi \right )
\left (f^2 + g^2 \right ) -
\kappa\frac{e^{-\nu}}{r\sqrt\Delta}fg -
\frac{\kappa}{2}m\frac{e^{-\nu}}{\sqrt\Delta}\left (f^2 - g^2\right ) ,
\label{ex12a}\\
r^2\Delta \phi '' & = & - 8\pi e
\left (f^2 + g^2 \right ) -
\left (
2r\Delta - r^2\Delta \nu '
 \right )\phi '
\label{ex13a}
\end{eqnarray}
where $\kappa$ is some constant.
This equations system was investigated in \cite{finster99} and
result is the following. A particle-like solution exists which
has the following expansions near $r = 0$
\begin{eqnarray}
f(r) & = & f_1r + {\cal O}(r^2),
\quad
g(r) = {\cal O}(r^2) ,
\label{ex11}\\
\Delta (r) & = & 1 + {\cal O}(r^2),
\quad
\nu(r) = {\cal O}(r^2) ,
\quad
\phi (r) = {\cal O}(r^2)
\label{ex12}
\end{eqnarray}
and the following asymptotical behaviour
\begin{eqnarray}
\Delta (r) & \approx & 1 - \frac{2m_\infty}{r} +
\frac{(2e_\infty)^2}{r^2}, \quad \nu (r) \approx const ,
\label{ex13}\\ \phi (r) & \approx & \frac{2e_\infty}{r} ,
\label{ex14}\\ f & \approx & f_0e^{-\alpha r}, \quad g \approx
g_0e^{-\alpha r}, \quad \frac{f_0}{g_0} = \sqrt{\frac{m_\infty +
\omega}{m_\infty - \omega}}, \quad \alpha ^2 = m_\infty^2 - \omega
^2. \label{ex15}
\end{eqnarray}
where $m_\infty$ is the mass for the observer at infinity and
$2e_\infty$ is the charge of this solution. Our interpretation of
this solution is presented on the Fig.(\ref{fig4}).
The solution exists for $(|e_\infty|/m_\infty)>1$ and
$(|e_\infty|/m_\infty)<1$\footnote{it depends on the mass $m$} but
for us is essential the first case with $(|e_\infty|/m_\infty)>1$.
In this case the classical Einstein-Maxwell theory leads to the
``naked'' singularity. The presence of the spacetime foam
drastically changes this result: \textit{the appearance of the
VWH's can prevent the formation of the ``naked'' singularuty in
the Reissner-Nordstr\"om solution with $|e|/m > 1$}.

\section{Conclusions}

Thus our model is based on the following assumptions:
\begin{itemize}
\item
in basic the quantum (virtual) WHs of the spacetime foam have the
5D throat.
\item
each quantum WH has a spin-like structure,
\item
the spacetime foam effectively can be described with the help of a
spinor field.
\end{itemize}
As the consequence we have the result that the strong electric
field separates the VWH's of the spacetime foam by such a way that
they can prevent a singularity in the Reissner-Nordstr\"om
solution (with $|e|/m>1$) on account of the formation of VWHs.
\par
In the spirit of the Einstein idea that the right-hand of
gravitational equations should be zero
we can note that this model of the WH with quantum throat is
the vacuum model since the gauge fields can be considered
as components of the metric in some multidimensional
Kaluza-Klein gravity.

\section{Acknowledgment}

I am grateful for financial support
from the Georg Forster Research Fellowship from the Alexander
von Humboldt Foundation and H.-J. Schmidt for an invitation to
Potsdam University.


\end{document}